\documentclass[10pt]{article}
\usepackage[OE]{express}
\usepackage[utf8]{inputenc}
\usepackage{booktabs}

\begin{document}
\title{Studying fundamental limit of optical fiber links to $10^{-21}$ level}

\author{Dan Xu,\authormark{1, *} Won-Kyu Lee,\authormark{1,3,4} Fabio Stefani,\authormark{1} Olivier Lopez,\authormark{2} Anne Amy-Klein,\authormark{2} and Paul-Eric Pottie\authormark{1}}

\address{\authormark{1}LNE-SYRTE, Observatoire de Paris, PSL Research University, CNRS, Sorbonne Universit\'{e}s, UPMC, 61 Avenue de l'Observatoire, 75014 Paris, France}
\address{\authormark{2}Laboratoire de Physique des Lasers, Universit\'{e} Paris 13, CNRS, 99 Avenue Jean-Baptiste Cl\'{e}ment, 93430 Villetaneuse, France}
\address{\authormark{3}Korea Research Institute of Standards and Science, Daejeon 34113, South Korea}
\address{\authormark{4}University of Science and Technology, Yuseong, Daejeon 34113, South Korea}

\email{\authormark{*}Dan.Xu@obspm.fr} 



\begin{abstract}
We present an hybrid fiber link combining effective optical frequency transfer and evaluation of performances with a self-synchronized two-way comparison. It enables us to detect the round-trip fiber noise and each of the forward and backward one-way fiber noises simultaneously. The various signals acquired with this setup allow us to study quantitatively several properties of optical fiber links. We check the reciprocity of the accumulated noise forth and back over a bi-directional fiber to the level of $3.1(\pm 3.9)\times 10^{-20}$ based on a 160000s continuous data. We also analyze the noise correlation between two adjacent fibers and show the first experimental evidence of interferometric noise at very low Fourier frequency. We estimate redundantly and consistently the stability and accuracy of the transferred optical frequency over 43~km at $4\times 10^{-21}$ level after 16 days of integration and demonstrate that frequency comparison with instability as low as $8\times 10^{-18}$ would be achievable with uni-directional fibers in urban area.
\end{abstract}



\section{Introduction}

Optical fiber links have been developed rapidly for time dissemination and frequency transfer over the last decade\,\cite{Newbury2007,Forman2007}. They are the most efficient technique for the distant comparison of the rapidly progressing optical atomic clocks\,\cite{Nicholson2015, Huntemann2016, Ushima2015, Schioppo2017, Falke2014,Nemitz2016}. These impressive progresses in optical frequency metrology pave the way for a future redefinition of the second\,\cite{Riehle2015}.

Optical frequency transfer via fiber networks has been performed with ever-increasing performance and fiber lengths. It demonstrates outstanding capabilities to transfer optical frequency with instability below $10^{-15}$ to the low $10^{-20}$ level, even on long-haul fiber links up to 1840~km\,\cite{Raupach2015, Chiodo2015, Droste2013}. Meantime, different schemes of optical fiber links have been investigated: the Active Noise Compensation (ANC) method\,\cite{Ma1994} and the passive noise rejection methods called as Two-Way (TW)\,\cite{Calosso2014}. In ANC method, the propagation noise is measured after a round-trip and actively compensated with an opto-electronic setup. On the contrary, TW comparison uses passive noise rejection after post-processing. 

The TW comparison method can be divided further into two types: classical two-way (CTW) and local two-way (LTW)\,\cite{Bercy2014,Stefani2015,Lee2017}. In CTW method, optical frequency comparisons are obtained by post-processing the data acquired from two physically separated locations, and efforts need to be paid to synchronize the acquisition of the data at two ends of the link \,\cite{Calosso2014,Grotti2017,Tampellini2015}. In LTW method, two beat notes are recorded in either of the laboratories: in addition to the "classical" data, one can record another optical beat note between the laser and itself after a round-trip, which is the same as the one detected in an ANC setup. In this case, by processing the data acquired locally, either only at the local site or only at the remote site, the propagation noise can be rejected in real time. The advantage is that the data are self-synchronized and optimal fiber noise rejection is achieved without extra efforts.

We investigate here an hybrid architecture, using an ANC setup on one fiber and using an LTW setup on the other parallel fiber. Such a topology has been reported in\,\cite{Lee2017} and allow us to analyze the contribution of laser and interferometric noise to stabilities. It is also used for optical frequency comparison between SYRTE (France) and NPL (UK)\,\cite{Delva2017}. We present here an improved hybrid setup with the motivation to use the parallel fibers for effective frequency transfer for direct application in optical metrology and served as a test bed for studying fundamental limits of fiber links. It enables us to detect round-trip and one-way fiber noise, from which we can infer several perfectly synchronous observables and show greater reliability and robustness of the setup. It enables us to assess the performance of optical frequency transfer in real time, while allowing extremely accurate frequency comparison of the primary and secondary frequency standards between National Metrology Institutes (NMIs), and providing effective outputs for advanced molecular spectroscopy experiment\,\cite{Argence2015,Delva2017}. Moreover, for the first time to the best of our knowledge, we can check the reciprocity of the accumulated noise forth and back over a bidirectional fiber, and we can measure the noise floor of an effective frequency transfer setup {\it in situ}. With this hybrid setup, we can also evaluate the noise correlation between two adjacent fibers or between forth and back propagation in one fiber. 

\begin{figure}[!b]
\centering
\includegraphics[width=10.5cm]{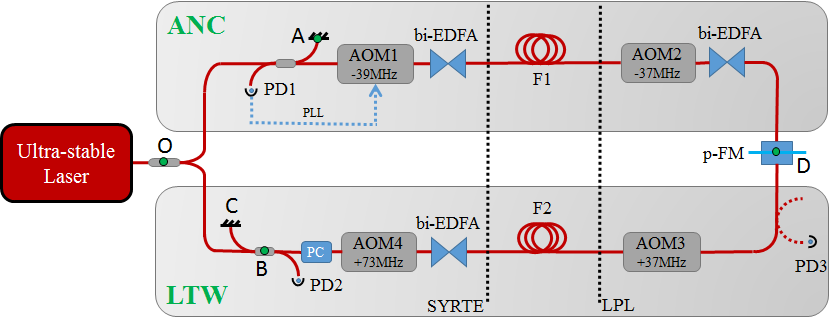}
\caption{Experimental scheme for the hybrid fiber links. AOM, acousto-optic modulator; PD, photodiode; PLL, phase-locked loop; bi-EDFA, bi-directional erbium-doped fiber amplifier; p-FM, partial Faraday mirror; PC, polarization controller.\label{Sketch}}
\end{figure}

\section{Experimental scheme}
The hybrid experimental scheme we implement for optical frequency transfer and comparison is shown in Fig.\,\ref{Sketch}. The two fibers, denoted by F1 and F2, are a pair of urban fibers of 43~km connecting two laboratories, SYRTE and LPL. An ANC setup is built on F1; an LTW setup is built on F2 as introduced in\,\cite{Lee2017}. The fibers are fed with one ultra-stable laser, de-drifted by referencing it to an H-Maser using an optical frequency comb. It is split into two parts: one part is injected into F1 (see upper part of Fig.\,\ref{Sketch}); the other is injected into F2, (see lower part of Fig.\,\ref{Sketch}). Two couplers at the local site are used to build the strongly unbalanced Michelson interferometers required for noise compensation. The Michelson interferometers are housed in an aluminum box enclosed in a polyurethane foam box for thermal and acoustic isolation and passive stabilization. Three bi-directional erbium-doped fiber amplifiers (bi-EDFAs) are used to compensate the transmission loss of the ultra-stable laser. A partial Faraday mirror (p-FM) is used at LPL site, which makes a perfect fiber-length match in the remote setup as the uncommon fiber length is zero\,\cite{Lee2017}. It enables the implementation of a simple but robust and temperature-insensitive interferometer at the end-user site. The 50/50 coupler and PD3 represented in dashed line in Fig.\,\ref{Sketch} lead to 6 dB additional losses. They were used in earlier work, but not effectively used in this experiment. The beat notes both on PD1 and PD2 at SYRTE, after detection, electronic amplification and filtering, are simultaneously recorded by two dead-time free frequency counters (K+K FXE) operated in $\Pi$-type and $\Lambda$-type with a gate time of 1~s. 

Based on the above hybrid scheme, we demonstrate two approaches of two-way frequency comparison. The first one uses two parallel fibers from the same cable, F1 and F2, each fiber transmitting the light in a single direction, referred to as two-way uni-directional (TWU). This approach enables us to estimate the noise correlation between the two adjacent fibers, F1 and F2. It opens the way to frequency comparison over telecommunication network with minimal modification of the network backbone. In addition, we realize a second approach which uses a single fiber, F2, through which the light propagates in both directions, referred to as two-way bi-directional (TWB). The TWB approach enables us to detect round-trip and one-way fiber noises and also check the reciprocity of the accumulated noise forth and back on F2.

\begin{figure}[!b]
\centering
\includegraphics[width=11.5cm]{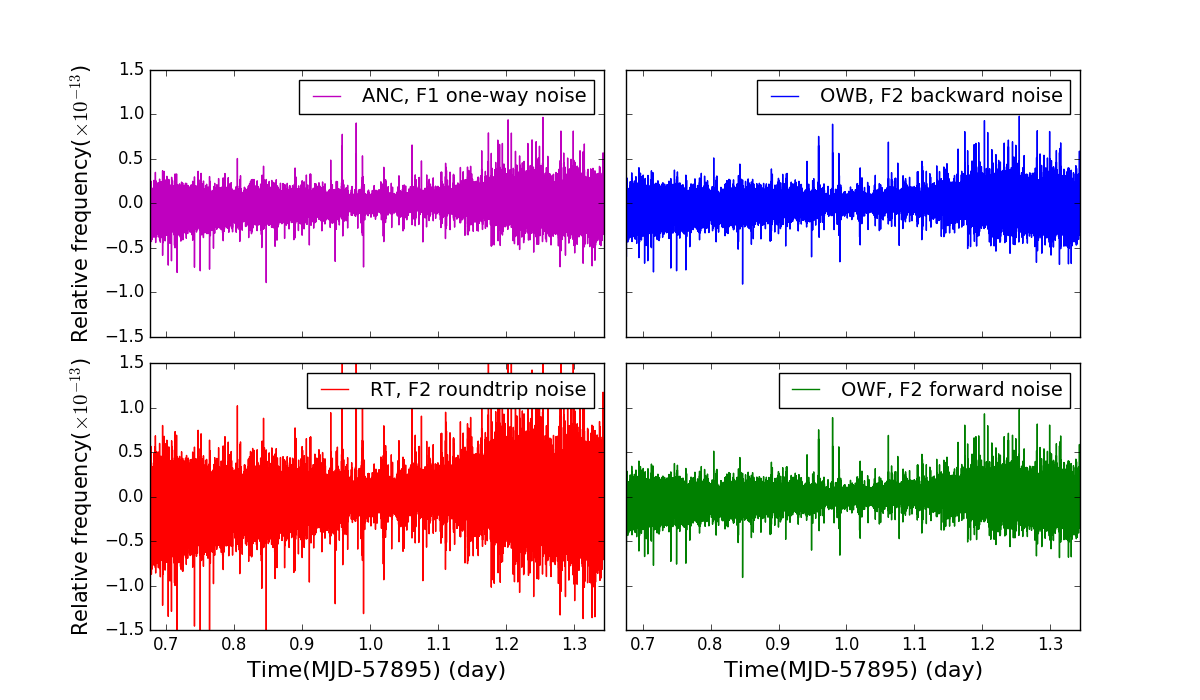}
\caption{Measurements of frequency data acquired only at SYRTE: half the round-trip noise of F1, round-trip noise, forward and backward one-way noise of F2 ($\Lambda$-counting).\label{RawTrace}}
\end{figure} 

\section{Experimental results}
At SYRTE, we obtain four beat notes: one on PD1 and three on PD2. On PD1, one records the beat note between the laser beam reflected by FM labeled A and by p-FM labeled D, shown in Fig.\,\ref{Sketch} as green filled circles. This signal exhibits the round-trip propagation noise on F1. It is detected, tracked and divided to apply an active noise compensation on AOM1. The correction of half the round-trip fiber noise on F1 is almost equal to one-way noise. We label this correction ANC (for Active Noise Compensation signal). After active noise compensation, the optical phase at A is then copied to the remote p-FM labeled D. On PD2, we obtain three beat notes: 1) the beat note between the light and itself after a round-trip on F2, denoted as round-trip term (RT); 2) the beat note between the local light and the virtual ultra-stable laser (the point D on p-FM), which is transferred backward on F2. We denote it the one-way backward term (OWB); 3) the beat note between the local light transferred forward on F2, reflected by p-FM, then transferred backward on F2, and the virtual ultra-stable laser transferred backward on F2. This beat note should be detected at LPL, but due to our experimental arrangement, thanks to the p-FM, the beat note is actually detected at SYRTE. We denote this beat note here one-way forward term (OWF). This term is weaker by about 30~dB. By using 2 bi-EDFAs at SYRTE and 1 bi-EDFA at LPL, OWF can be detected at SYRTE. Using a double-down-conversion stage and narrow RF filtering with bandwidth less than 1~MHz, and tracking and dividing the beat notes, stable counting with almost no cycle slips is achieved.
The four beat notes contain abundant information about the two parallel fibers' noise: the one-way noise of F1 (ANC), the round-trip noise of F2 (RT), the one-way forward noise of F2 (OWF), the one-way backward noise of F2 (OWB). Their relative frequency fluctuations with $\Lambda$-data are shown in Fig.\,\ref{RawTrace}. As expected, the one-way noises of F1 (ANC) and F2 (OWB, OWF), are almost at the same level. The round-trip noise on F2 (RT) is twice the other terms.

\begin{figure}[!t]
\centering
\includegraphics[width=10.5cm]{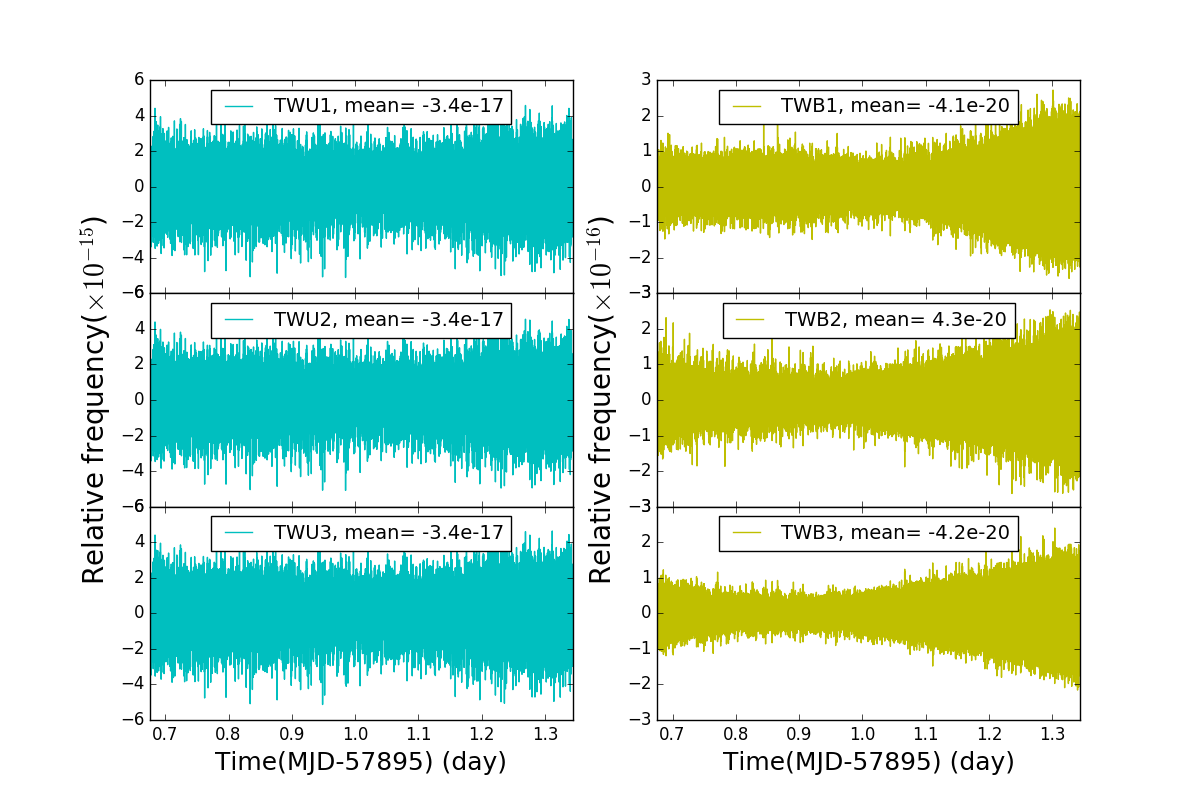}
\caption{Measurement of three sets of TWU and three sets of TWB by combining the four data sets: ANC, RT, OWB, OWF ($\Lambda$-counting).\label{TWUandTWB}}
\end{figure} 

Based on the above four basic measurements acquired only at SYRTE, we can obtain 7 observables. Three observables which arise from two separated fibers in the same cable used for the uplink and downlink (F1 and F2), are given by
\begin{equation}
  \begin{cases}
  \phi_{\text{TWU1}}=\phi_{\text{ANC}}-\phi_{\text{RT}}/2\\
  \phi_{\text{TWU2}}=\phi_{\text{ANC}}-\phi_{\text{OWB}}.\\
  \phi_{\text{TWU3}}=\phi_{\text{ANC}}-\phi_{\text{OWF}}
  \end{cases}
\end{equation}
These uni-directional observables are of the same nature as the uni-directional data reported in\,\cite{Bercy2014,Lee2017,Williams2008}, or similar architecture in DWDM network as reported in\,\cite{Turza2017}. We interpret them as a virtual uni-directional LTW setup where the forward and backward signals propagate in two different(uni-directional) fibers, thus the observables TWUs give the difference between the one-way fiber noise of F1 and half the round-trip, forward and backward one-way fiber noise of F2, respectively. As shown in Fig.\,\ref{TWUandTWB}, TWUs behave consistently which give a first indication of high noise correlation between the two adjacent fibers. These correlations have been estimated and differ from 1 at the level of $10^{-4}$ with $\Lambda$-counting phase data. It indicates that the phase noises between two fibers are indeed highly correlated. The main limiting factors of TWUs arise from the remaining uncorrelated fiber noises. 
  
The three following observables which arise from one fiber (F2), are given by
\begin{equation}
  \begin{cases}
  \phi_{\text{TWB1}}=\phi_{\text{OWB}}-\phi_{\text{RT}}/2\\
  \phi_{\text{TWB2}}=\phi_{\text{OWF}}-\phi_{\text{RT}}/2.\\ 
  \phi_{\text{TWB3}}=(\phi_{\text{OWB}}-\phi_{\text{OWF}})/2 
  \end{cases}
\end{equation}
We interpret them as a fully bi-directional LTW setup. In our previous setup, TWB1 was detected at SYRTE and TWB2 at LPL; TWB3 was obtained by post-processing the data between SYRTE and LPL\,\cite{Lee2017}. In this experimental arrangement, TWB1, TWB2 and TWB3 are obtained at SYRTE, and all of them are self-synchronized. Note that TWB3 is a classical two-way term although  no data exchange is needed. The correlations between half the round-trip, forward and backward one-way fiber noise of F2 were also evaluated and differ from 1 at the level of around $10^{-10}$ with $\Lambda$-counting phase data. The rejection of fiber noise can be almost perfect after some integration, and thus the dominating noise for TWBs is interferometric noise.

In addition, we introduce here a new two-way noise floor(TWNF) observable: 
\begin{equation}
  \phi_{\text{TWNF}}=(\phi_{\text{OWB}}+\phi_{\text{OWF}}-\phi_{\text{RT}})/2 .
\end{equation}
With this definition, if the noises forth and back are equal, the observable should converge to zero. The interferometric noise of OWB, OWF and RT exactly compensate themselves (see Appendix for details), then we expect that the TWNF exhibits interferometric noise free behavior. Such an interferometric noise here specially arise from the fibers of the local interferometers used for the beatnotes detection. As it is the sum of three independent signals, it contains also the quadratic sum of the noise floors of the three signals. To the best of our knowledge, this is the first time that the noise floor of an effective frequency transfer setup can be measured {\it in situ}.

In the following section, we present first a detailed analysis of the phase evolution and frequency stability for the bi-directional observables. The accuracy and long-term behavior of uni-directional and bi-directional observables are presented in section 5.

\section{Analysis and discussion of the bi-directional observables}
\subsection{Phase evolution and Fourier analysis} 

\begin{figure}[!t]
\centering
\includegraphics[width=13cm]{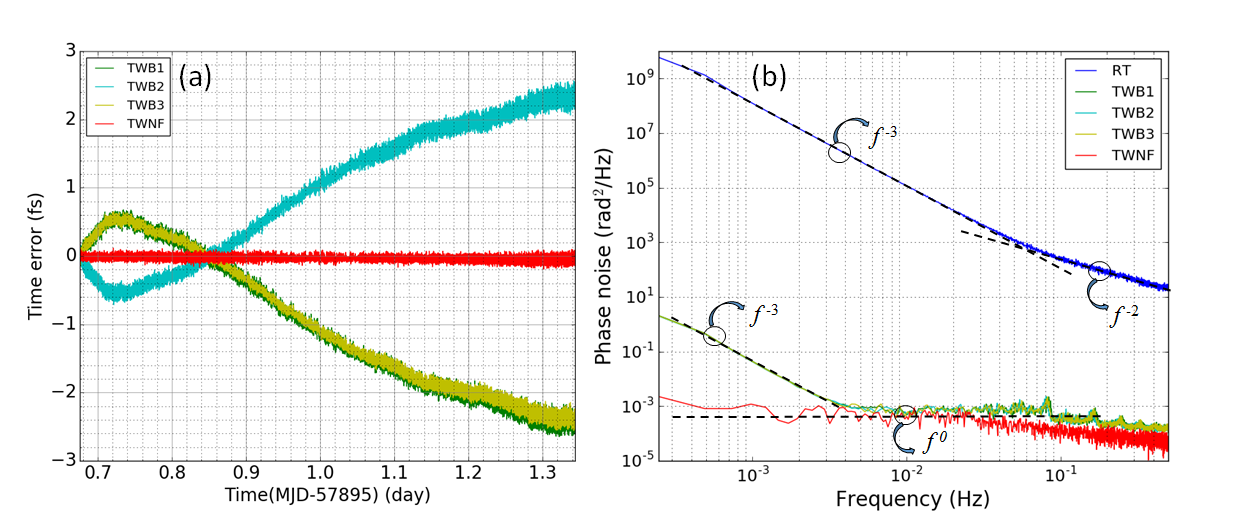}
\caption{(a) Phase evolution of the bi-directional LTW observables: TWBs and TWNF, (b) phase noise power spectral densities(PSD) of the roundtrip noise on F2 and the residual phase noise by bi-directional LTW method. Note that three PSDs corresponding to TWB1, TWB2 and TWB3 are overlaid on each other.
\label{PhaseTrace}}
\end{figure}

In order to provide a better insight of the differences between these TWBs and TWNF observables, we present their phase evolution in Fig.\,\ref{PhaseTrace}(a). Phase error is actually converted into time error and expressed in fs. The three TWBs show the same phase evolution, about 3~fs over one day of measurement. Obviously, TWB1 and TWB2 have opposite sign, TWB1 and TWB3 coincide,  which are expected from our definitions. Knowing the temperature evolution of the experimental room, the temperature sensitivity of optical fibers and the length's mismatch of our interferometric setup from previous work, the phase evolution of TWBs can be attributed to interferometric noise\,\cite{Lee2017}.

By contrast, TWNF has a phase evolution less than 0.2~fs, as expected. Fig.\,\ref{PhaseTrace}(a) illustrates that the phase wandering seen with TWBs are almost erased in TWNF. This is an evidence that TWNF is interferometric-noise free. This is a new situation compared to the noise floor measurements reported elsewhere previously. Usually noise floor is estimated by optically short-cutting the link with a short fiber patch cord, and setting the detection parameter nearly the same as those in real experimental conditions. In that case, the bandwidth of the servo is inherently changed, as the propagation time delay is modified. The interferometric noise of the setup is kept unmodified. However, in this new situation, it is the opposite: the detection parameters are the very same as the ones of the running link : the propagation time delay is not modified, power levels are the same, but the interferometric noise is erased from the data set combination.

To study further the properties of the TWBs and TWNF, we derive the power spectral density (PSD) by calculating the fast Fourier transform of the auto-correlation of their phase noise\,\cite{Stefani2015}, as shown in Fig.\,\ref{PhaseTrace}(b). We find out that in our hybrid fiber link, from 0.5~mHz to about 80~mHz, the PSD of RT falls off as a flicker frequency noise $f^{-3}$, where $f$ is the Fourier frequency. Then the RT noise changes to white frequency noise ($f^{-2}$) above 80~mHz. It is interesting to note that the PSD of TWBs also fall off as $f^{-3}$ from 0.5~mHz to 5~mHz. It indicates that the TWBs residual noise are originating from fiber noise arising from non-compensated fiber segments, thus the interferometers' noises dominate the TWBs noise at low frequencies. For higher Fourier frequency, approximately white phase noises are achieved for TWBs at about $10^{-3}$~rad$^{2}$/Hz, indicating that the phase noise is canceled within the bandwidth. The PSD of TWNF shows only white phase noise signature over almost the full frequency range. The constant slope  even at very low Fourier frequency confirms that TWNF is interferometric-noise free. More informations are detailed in Appendix.

\subsection{Frequency stability} 

\begin{figure}[!t]
\centering
\includegraphics[width=8.5cm]{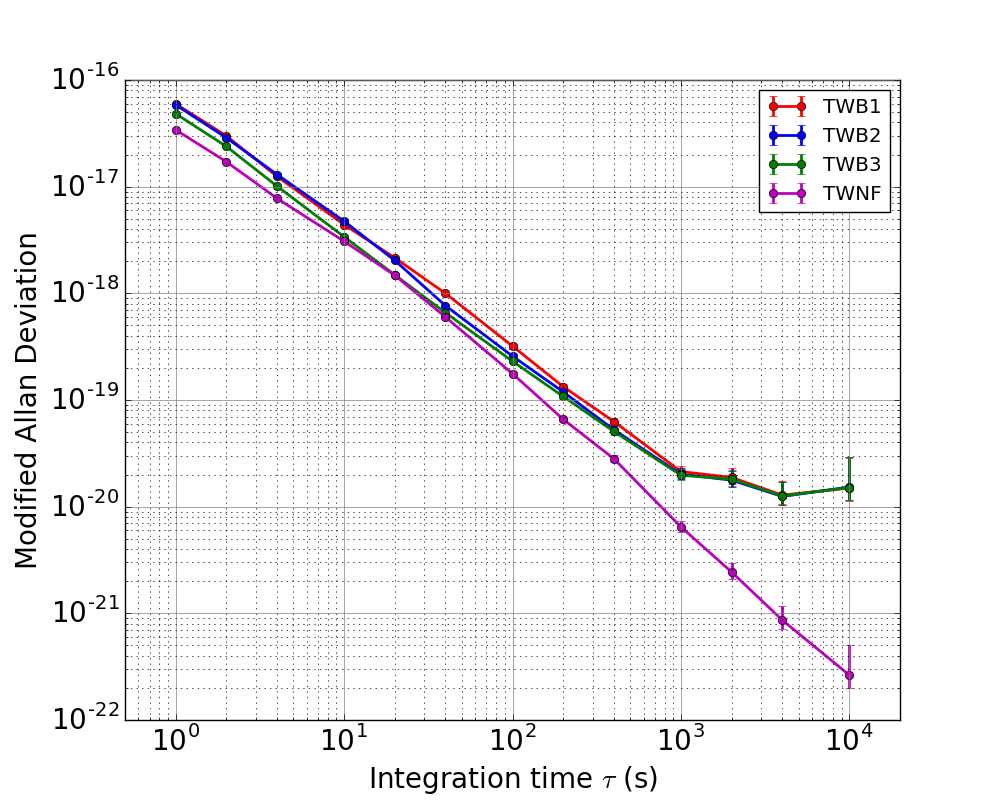}
\caption{Modified Allan deviation(MDEV) of 3 bi-directional LTW observables TWB1, TWB2, TWB3, as well as TWNF ($\Lambda$-counting). \label{MDEV}}
\end{figure}  

The fractional frequency instabilities of TWBs and TWNF are plotted in Fig.\,\ref{MDEV} as modified Allan deviation (MDEV) calculated from the $\Lambda$-data\,\cite{Dawkins2007}. The stability of TWBs is as low as $5\times 10^{-17}$ at 1~s integration time. The stability of TWB3 is lower than that of TWB1 and TWB2, which could arise from the different detection noises. We expect indeed that the detection noise contribution is higher for TWB1 and TWB2 than for TWB3, for which the noise contribution of the backward-F2 and forward-F2 beatnotes are both divided by two, following Eq.7 in Appendix. At long integration time, the measurements reach the interferometric noise floor, and the three curves of TWBs perfectly merge themselves. The TWBs reach relative frequency stabilities of $1.5\times 10^{-20}$ at 10\,000s integration time. At the longer integration time, the MDEV of TWNF deviates from TWBs and keeps decreasing with a slope of $-\frac{3}{2}$. TWNF reaches $2.6\times 10^{-22}$ at 10\,000~s, which further indicates that TWNF is insensitive to diurnal noise bump due to temperature variations.

\begin{figure}[!t]
\centering
\includegraphics[width=8.5cm]{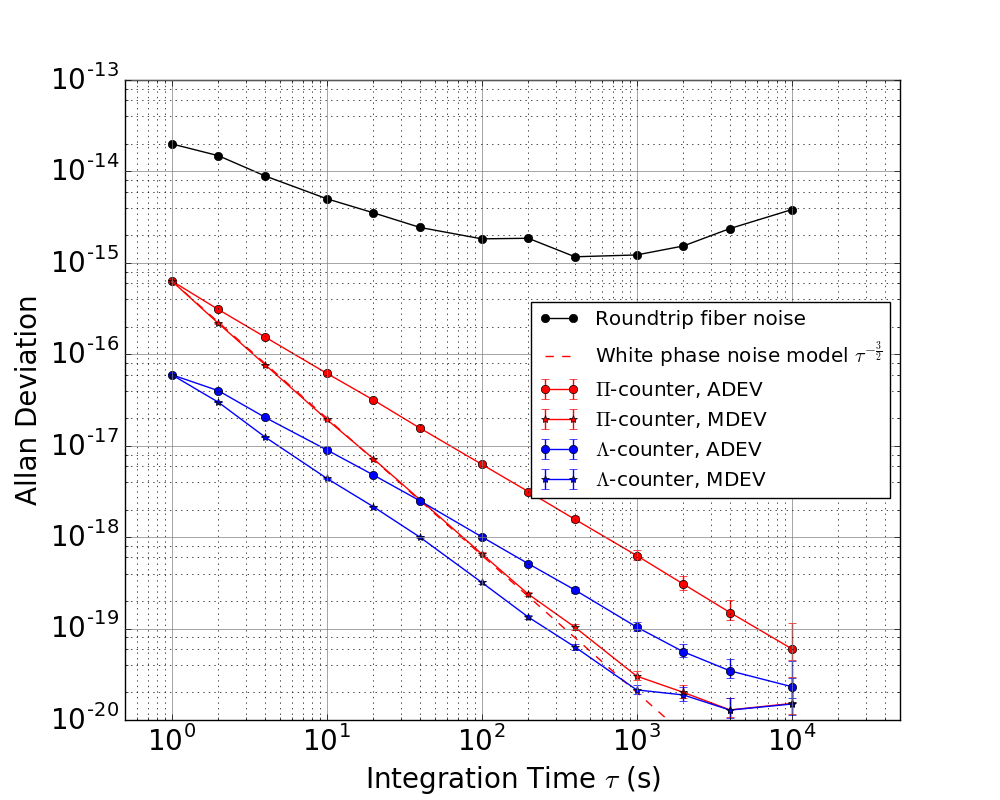}
\caption{Fractional frequency stabilities of TWB1 in terms of overlapping Allan deviation (ADEV) and modified Allan deviation (MDEV) with $\Pi$-type counter and $\Lambda$-type counter.\label{ADEV_TWB}}
\end{figure}	 

Figure\,\ref{ADEV_TWB} shows the fractional frequency instabilities of bi-directional LTW in terms of overlapping Allan deviation (ADEV) and modified Allan deviation (MDEV). The free running round-trip fiber noise of F2 is shown as a black line. We show only TWB1 for the sake of clarity, but similar data were obtained for TWB2 and TWB3. The ADEV and MDEV of TWB1 that are calculated from the $\Pi$-data start from $6.3\times 10^{-16}$ at 1~s integration time and decrease with the slope of $-1$ and $-\frac{3}{2}$, respectively. The ADEV and MDEV of TWB1 from the $\Lambda$-data start from $5.9\times 10^{-17}$ at 1~s and decrease to $2.1\times 10^{-20}$ and $1.5\times 10^{-20}$ at 10\,000s, respectively. The MDEVs of the $\Pi$-data and $\Lambda$-data converge to the same value and show good agreement. The ADEVs of the $\Pi$-data and $\Lambda$-data show a difference compatible with the statistical error bars. Actually we check that ADEVs converge to the same value at longer integration time, as seen in Table\,\ref{Table:Accuracy}. Compared to the results shown in\,\cite{Lee2017}, we find excellent agreement and consistent behavior over 2 years of operation. The ratio between the $\Lambda$-data and $\Pi$-data at 1~s integration time is 10.6. Using K+K counter, the maximum expected ratio for pure white phase noise is approx. 31. The $\tau^{-1}$ behavior of the MDEV using $\Lambda$-data and the noise floor measurement given by TWNF show that we are limited by the noise floor of the setup. We check this assumption by replacing the fiber link by a short patch fiber and setting an equivalent remote end (AOMs and p-FM) at SYRTE. We observe the same noise floor limitation from 1~s to 100~s integration time.       

\section{Uni-directional and bi-directional data over days}
\subsection{Accuracy and reciprocity}
We evaluate the accuracy of the frequency comparison of the hybrid setup. We calculate the arithmetic mean of the difference between the expected and the measured beat frequency of the relevant observables TWUs and TWBs introduced above. The statistical uncertainty is estimated by the instability of the data set at a well-considered averaging time\,\cite{Raupach2015,Lee2017,Lee2010,Benkler2015}. The effects of missing data in comparisons were explored in the case of GPS and satellite two-ways, but not in the context of coherent fiber links\,\cite{Sesia2016,Defraigne2008}. The question is not trivial, and out of the scope of this paper. So we use here a subset of 160\,000 s (2 days) data without any cycle slip, recorded with 1~s gate time. The results for this cycle-slip-free 160\,000 s data are presented in Table\,\ref{Table:Accuracy}, with the mean value, ADEV and MDEV at 40\,000\  s of integration time, with $\Pi$-data and $\Lambda$-data. 

The three TWU and three TWB are consistent, which show the reliability of the setup and give a greater confidence into the quality of the data set. We observe the convergence of ADEVs using $\Pi$-data and $\Lambda$-data. The MDEVs converge also to identical values. No systematic shifts are observed for all the observables within the statistical uncertainty at 1-$\sigma$ level.   

For uni-directional LTW, we obtain mean value of these relative frequencies of $2.2\times 10^{-17}$, with a statistical uncertainty of $1.0\times 10^{-17}$ and $7.0\times 10^{-18}$ given by the long-term ADEV and MDEV at 40\,000 s ($\Pi$-data and $\Lambda$-data), respectively. For bi-directional LTW, the mean values are $\sim3.0\times 10^{-20}$, with a statistical uncertainty of $3.9\times 10^{-20}$ and $2.8\times 10^{-20}$ given by the long-term ADEV and MDEV at 40\,000 s ($\Pi$-data and $\Lambda$-data), respectively. As a conclusion, the accuracy of our setup is limited by the statistical uncertainty contribution. We consider here then conservatively the worse case, to be $3.9\times 10^{-20}$.

In particular, the TWB3 can be interpreted as a test of reciprocity using in field fibers between the forward and backward paths, at the level of $3.1(\pm 3.9) \times 10^{-20}$. It is limited by the interferometric noise at SYRTE. The effect of a periodic perturbation can be also a problem at such high level of accuracy. Improvement of the setup and a better data processing should allow us to improve the accuracy of this first test of reciprocity.   
\begin{table}[!h]
\caption{Accuracy of the frequency comparisons using the hybrid architecture\label{Table:Accuracy}}
\centering
 \begin{tabular}{l ccc ccc}
 \toprule
   Quantity & TWU1 &  TWU2 &  TWU3 &  TWB1 &  TWB2 & TWB3 \\
  		   & \multicolumn{3}{c}{$\times 10^{-17}$} & \multicolumn{3}{c}{$\times 10^{-20}$}\\
  \hline\addlinespace
  ~~~~~~~~~~Mean &  $2.2$ &  $2.2$ &  $2.2$ &  $2.9$ &  $3.2$ & $3.1$ \\
  \raisebox{1ex}{$\Pi$}
  ~~~~~~ADEV at 40~000s &  $1.0$ &  $1.0$ &  $1.0$ &  $4.0$ &  $4.0$ & $3.9$ \\[0.5ex]  
  ~~~~~~~~~~MDEV at 40~000s &  $0.7$ &  $0.7$ &  $0.7$ &  $2.8$ &  $2.8$ & $2.8$ \\
  \hline\addlinespace
 ~~~~~~~~~~Mean &  $2.2$ &  $2.2$ &  $2.2$ &  $3.0$ &  $2.9$ & $2.9$\\
  \raisebox{1ex}{$\Lambda$}
  ~~~~~~ADEV at 40~000s &  $1.0$ &  $1.0$ &  $1.0$ &  $3.9$ &  $3.9$ & $3.9$ \\[0.5ex]
 ~~~~~~~~~~MDEV at 40~000s &  $0.7$ &  $0.7$ &  $0.7$ &  $2.8$ &  $2.8$ & $2.8$ \\
 \bottomrule
 \end{tabular}
\end{table}

\subsection{Behavior over 16 days} 
\begin{figure}[!b]
\centering
\includegraphics[width=13cm]{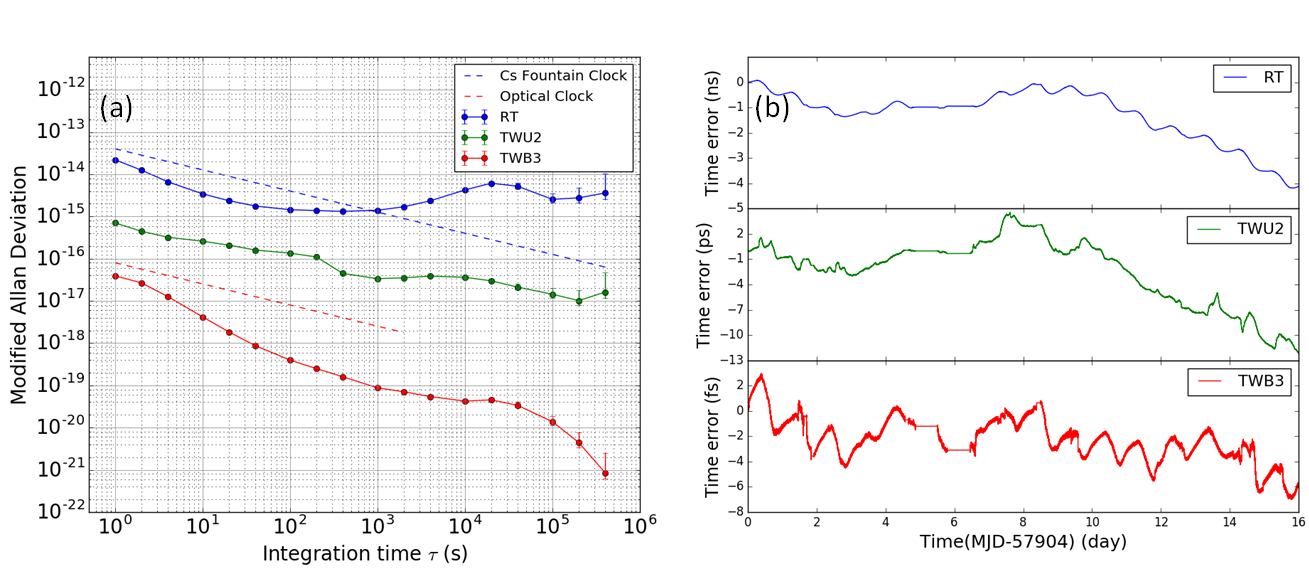}
\caption{(a) Long-term fractional frequency instabilities and (b) phase evolution of a uni-directional and a bi-directional two-way scheme($\Lambda$-counting). \label{ADEVLong}}
\end{figure}

The hybrid setup runs continuously for one month, as a part of an international clock comparison between NPL (UK), SYRTE (France) and PTB (Germany). We show in Fig.\,\ref{ADEVLong} the stability and phase evolution of the observables of the hybrid fiber link over 16 days. The interruptions seen on the plot at MJD = 57904 + 5 and +6 are not due to the setup, but to an interruption of the ultra-stable light (no input signal). Except for the interruptions, the uptime over 14 days is estimated conservatively to 96.5$\%$. RT has a phase evolution less than 4~ns; TWU2 is at the level of ~ps; TWB3 is at the level of ~fs. The MDEV of TWU2 starts from $7.0\times 10^{-16}$ at 1~s and decreases to $1.3\times 10^{-17}$ at 400~000s integration time. The MDEV of TWB3 is $3.9\times 10^{-17}$ at 1~s and reaches as low as $8.5\times 10^{-22}$ at 400~000s integration time. It shows the reliability of the hybrid setup in the long-term. The stability curve of TWB3 shows a plateau from $10^{4}$\, s to $4\times 10^{4}$\, s integration time, and then a rapid decay with $\tau^{-2}$, indicating a periodic perturbation with a time constant of approximately 40\,000 s, which we attribute to the diurnal temperature fluctuations of the fiber, which is also evidenced by the stability curve of RT [blue dots in Fig.\,\ref{ADEVLong}(a)]\,\cite{Raupach2014}.

The typical stability of Cs Fountain Clock is shown as blue dashed line, taken as $4\times 10^{-14}/\sqrt\tau$; the red dashed line shows the stability of one of the best optical clocks, taken as $8\times 10^{-17}/\sqrt\tau$, as reported for instance in\,\cite{Schioppo2017}. For achieving optical frequency comparison with the best accuracy, it is mandatory to use fully bi-directional setup. Moreover, the data acquired with our hybrid setup confirm the first results reported in [14], and show the ability of uni-directional setup to compare the best microwave cold atoms frequency standards, at least at a regional scale.

We use the data set of 16 continuous days for accuracy analysis. For TWU2, the mean offset frequency of $8.8\times 10^{-18}$ shows nearly no deviation within the statistical uncertainty of $1.3\times 10^{-17}$. For TWB3, we obtain a mean offset frequency of $4.2\times 10^{-21}$, that deviates from its statistical uncertainty of $8.5\times 10^{-22}$. This frequency shift could arise from subtle technical issues or from the limited time of observation (as compared to the 1-day periodic perturbation presented in this data set). The effect is not explained so far with simple models and need detailed analysis over very long run, and will be the object of future studies. Anyway, although a conclusion cannot be made about the origin of the effect, it is not a limiting factor for the accuracy of optical clock comparison by {\it at least} 2 orders of magnitude. 

\section{Conclusion}
We present fully new data and data analysis for the $2\times 43$~km hybrid fiber link. We demonstrate an hybrid setup suitable for the study of fundamental limits of fiber links, while being a very effective link for optical frequency dissemination and optical frequency comparison. We show that the hybrid setup evaluates uni-directional LTW abilities to reach instability below $10^{-17}$ over 43~km of an urban network. The hybrid setup we have implemented enables us to transfer frequency with a relative frequency stability as low as $3.9\times 10^{-17}$ at 1~s integration time and below $10^{-21}$ over 16 days. Combined with another hybrid links between LPL and NPL, this hybrid setup is already a part of an "optical fiber links chain" linking NPL, SYRTE and PTB\,\cite{OFTEN}. With this work, we are showing that far from being competitive, the combination of two techniques can even lead to improve efficiency, redundancy and overall performances. 

\section*{Funding}
We acknowledge funding support from the Agence Nationale de la Recherche (Labex First-TF ANR-10-LABX-48-01, Equipex REFIMEVE+ANR-11-EQPX-0039), Conseil R\'{e}gional Ile-de-France (DIM Nano'K), CNRS with Action Sp\'{e}cifique GRAM, the European Metrology Research Programme (EMRP) in the Joint Research Projects SIB02 (NEAT-FT) and the European Metrology Programme for Innovation and Research (EMPIR) in project 15SIB05 (OFTEN). The EMRP and EMPIR are jointly funded by the EMRP/EMPIR participating countries within EURAMET and the European Union. W.-K. Lee was supported partly by the Korea Research Institute of Standards and Science under the project "Research on Time and Space Measurements", Grant No.~16011007, and also partly supported by the R$\&$D Convergence Program of NST (National Research Council of Science and Technology) of Republic of Korea (Grant No.~CAP-15-08-KRISS).

\section*{Appendix\label{AppendA}}
Here, we analyze in detail the phase noise in the hybrid setup. According to the noise sources, the total phase noise of the beat note can be given by the sum of four terms splitting the phase wandering into the laser term, the RF term, the interferometric term and the fiber term:
\begin{equation}
  \phi=\phi_{\text{laser}}+\phi_{\text{RF}}+\phi_{\text{inter}}+\phi_{\text{fiber}} \label{noise}
\end{equation}
Since the same laser is used to inject the two fibers, and its frequency is almost constant, the first term can be neglected and we state $\phi_{\text{laser}}=0$. By phase-locking the PLL reference synthesizer and the frequency counters to a common 10~MHz reference signal, the RF contribution to phase can be neglected, and we state $\phi_{\text{RF}}=0$. The interferometric noise is due to the length mismatch in the local interferometers (a perfect match of the fiber length in the remote interferometer is accomplished by the use of a partial Faraday mirror at the remote setup)\,\cite{Lee2017}. As the interferometers are isolated passively in a box with time constant which is much greater than the light propagation time in the fiber, we write the interferometric term as $\phi_{\text{inter}}(t)$. Thus for the four basic beat notes, the two dominant phase noise are the fiber link noise and interferometers' noise

\begin{eqnarray}
  \left\{
  \begin{aligned}
    \phi_{\text{ANC}}(t)&= \phi_{\text{inter-ANC}}(t)+\frac{1}{2}\phi_{\text{RT-F1}}(t)\\
    \phi_{\text{RT}}(t)&=\phi_{\text{inter-RT}}(t)+\phi_{\text{RT-F2}}(t)\\
    \phi_{\text{OWB}}(t)&=\phi_{\text{inter-OWB}}(t)+\phi_{\text{backward-F2}}(t)\\
    \phi_{\text{OWF}}(t)&=\phi_{\text{inter-OWF}}(t)+\phi_{\text{forward-F2}}(t-\tau)
  \end{aligned}
  \right.
\end{eqnarray}
where $\tau$ is the light propagation delay between the local site and the remote site, $\phi_{\text{inter-ANC}}$ is the interferometric noise in the ANC setup, $\phi_{\text{RT-F1}}$ is the roundtrip propagation noise on F1, and other notations have the similar definitions.
Then the TWUs can be expressed as

\begin{eqnarray}
  \left\{
  \begin{aligned}
  \phi_{\text{TWU1}}(t)&=\phi_{\text{inter-TWU1}}(t)+\frac{1}{2}\phi_{\text{RT-F1}}(t)-\frac{1}{2}\phi_{\text{RT-F2}}(t)\\
  \phi_{\text{TWU2}}(t)&=\phi_{\text{inter-TWU2}}(t)+\frac{1}{2}\phi_{\text{RT-F1}}(t)-\phi_{\text{backward-F2}}(t)\\
  \phi_{\text{TWU3}}(t)&=\phi_{\text{inter-TWU3}}(t)+\frac{1}{2}\phi_{\text{RT-F1}}(t)-\phi_{\text{forward-F2}}(t-\tau)
  \end{aligned}
  \right.
\end{eqnarray}  

The TWBs can be expressed as

\begin{eqnarray}
  \left\{
  \begin{aligned}
  \phi_{\text{TWB1}}(t)&=\phi_{\text{inter-TWB1}}(t)+\phi_{\text{backward-F2}}(t)-\frac{1}{2}\phi_{\text{RT-F2}}(t)\\
  \phi_{\text{TWB2}}(t)&=\phi_{\text{inter-TWB2}}(t)+\phi_{\text{forward-F2}}(t-\tau)-\frac{1}{2}\phi_{\text{RT-F2}}(t)\\
  \phi_{\text{TWB3}}(t)&=\phi_{\text{inter-TWB3}}(t)+\frac{1}{2}[\phi_{\text{backward-F2}}(t)-\phi_{\text{forward-F2}}(t-\tau)]
  \end{aligned}
  \right.
\end{eqnarray}
It is demonstrated that the interferometric noises arise from the fiber sensitivity to external perturbations. Since acoustic and mechanic noises can be made negligible thanks to the well-designed isolation box, the so-called interferometric noise is mainly the result of long-term thermal effects. Inside the interferometers enclosure, the temperature is approximately homogeneous and we have
\begin{equation}
  \phi_{\text{inter}}=2\pi\nu\gamma\Delta T\delta L
\end{equation}
where $\gamma$ is a phase-temperature coefficient of silica fiber, which has a value of 37 fs/(K$\cdot$m) for an optical carrier at 194.4 THz and at 298 K, and $\Delta T$ is the temperature variation in the local interferometers. We omit here the time dependency for sake of clarity. For OWB, OWF and RT, the temperature variation is the same, but the fiber length mismatch is different. The local fiber length mismatch consists of the uncommon paths from the ultra-stable laser to two Michelson-like interferometers and the two short arms of the two interferometers. Following the approach developed in\,\cite{Lee2017}, the RT term will contain an interferometric noise proportional to the length 2$L_{\text{BC}}$; the OWB term contains an interferometric noise proportional to the optical length difference 2$L_{\text{BC}}+L_{\text{OA}}-L_{\text{OB}}$; the OWF term contains an interferometric noise only proportional to  the optical length difference $L_{\text{OB}}-L_{\text{OA}}$ (see Fig.\,\ref{Sketch} for the definition of point O, A, B and C). Thus the interferometric noise are given by 

\begin{eqnarray}
  \left\{
  \begin{aligned}
  \phi_{\text{inter-RT}}&=2\pi\nu\gamma\Delta T\cdot 2L_{\text{BC}}\\  
  \phi_{\text{inter-OWB}}&=2\pi\nu\gamma\Delta T\cdot(2L_{\text{BC}}+L_{\text{OA}}-L_{\text{OB}})\\
  \phi_{\text{inter-OWF}}&=2\pi\nu\gamma\Delta T\cdot(L_{\text{OB}}-L_{\text{OA}})
  \end{aligned}
  \right.
\end{eqnarray}
And the interferometric noise term for TWBs can be expressed as
\begin{equation}
  \begin{cases}
  \phi_{\text{inter-TWB1}}=2\pi\nu\gamma\Delta T\cdot(L_{\text{BC}}+L_{\text{OA}}-L_{\text{OB}})\\  
  \phi_{\text{inter-TWB2}}=2\pi\nu\gamma\Delta T\cdot(-L_{\text{BC}}-L_{\text{OA}}+L_{\text{OB}})\\
  \phi_{\text{inter-TWB3}}=2\pi\nu\gamma\Delta T\cdot(L_{\text{BC}}+L_{\text{OA}}-L_{\text{OB}})
  \end{cases}
\end{equation}
The phase evolutions for TWB1 and TWB2 are opposite, and those for TWB1 and TWB3 are the same, as shown in Fig.\,\ref{PhaseTrace}(a). We measured in\,\cite{Lee2017} that the length mismatch $L_{\text{BC}}+L_{\text{OA}}-L_{\text{OB}}\approx15$~cm. The phase evolution shown in Fig.\,\ref{ADEVLong}(b) is compatible with a temperature variation of 1.5 K, a temperature variation of the local interferometers that we actually confirm experimentally during the acquisition period. 

For the definition of TWNF, under the combination of OWB, OWF and RT, the interferometric noises compensate themselves, so
\begin{equation}
  \phi_{\text{inter-TWNF}}=0
\end{equation}
Showing that TWNF is interferometric-noise free. 

\section*{Acknowledgments}
The authors would like to thank Florian Frank for providing assistance and help in many experimental aspects.
\end{document}